\documentclass[10pt,twocolumn]{article}

\usepackage{arxiv}
\usepackage{times}
\usepackage{epsfig}
\usepackage{graphicx}
\usepackage{amsmath}
\usepackage{amssymb}
\usepackage[pagebackref=true,breaklinks=true,letterpaper=true,colorlinks,bookmarks=false]{hyperref}
\usepackage{booktabs}
\usepackage{multirow, makecell}

\usepackage{soul}
\usepackage{amsmath}
\usepackage{amssymb}
\usepackage{microtype}
\usepackage{wrapfig}

\def\figurePath{arxivimages/}
\def\myfigure#1#2{\begin{figure}[htb]\centering\includegraphics*[width = \linewidth]{\figurePath#1}\caption{#2}\label{fig:#1}\end{figure}}
\def\mycfigure#1#2{\begin{figure*}[t]\centering\includegraphics*[clip, width = \linewidth]{\figurePath#1}\caption{#2}\label{fig:#1}\end{figure*}}

\renewcommand{\eg}{e.\,g., }
\renewcommand{\ie}{i.\,e., }
\renewcommand{\etal}{et~al.\ }

\newcommand{\refSec}[1]{Sec.~\ref{sec:#1}}
\newcommand{\refFig}[1]{Fig.~\ref{fig:#1}}

\newcommand{\refTbl}[1]{Tbl.~\ref{tbl:#1}}

\soulregister\ref7
\soulregister\cite7
\soulregister\refFig7
\soulregister\cite7
\soulregister\ref7
\soulregister\pageref7
\soulregister\shortcite7
\soulregister\eg0
\soulregister\ie0
\soulregister\etal0
\soulregister\reftbl7

\DeclareGraphicsExtensions{.png,.jpg,.pdf,.ai,.psd}
\newcommand{\mysection}[2]{\section{#1}\label{sec:#2}}
\newcommand{\mysubsection}[2]{\subsection{#1}\label{sec:#2}}
\newcommand{\mysubsubsection}[2]{\subsubsection{#1}\label{sec:#2}}

\begin{document}

\title{Learning to Predict Image-based Rendering Artifacts\\with Respect to a Hidden Reference Image}

\author{
    \parbox{\textwidth}{
        \centering
        Mojtaba Bemana$^1$\ \ \ 		 
        Joachim Keinert$^2$\ \ \ 		 
        Karol Myszkowski$^1$\ \ \ 
        Michel B\"atz$^2$\ \ \ 
        Matthias Ziegler$^2$\ \ \ 
        Hans-Peter Seidel$^1$\ \ \ 
        Tobias Ritschel$^3$
    }
    \\
    \\
    \parbox{\textwidth}{
        \centering
        $^1$MPI Informatik\qquad
        $^2$Fraunhofer IIS\qquad			 
        $^3$University College London
    }
}

\maketitle

\begin{abstract}
Image metrics predict the perceived per-pixel difference between a reference image and its degraded (\eg re-rendered) version.
In several important applications, the reference image is not available and image metrics cannot be applied.
We devise a neural network architecture and training procedure that allows predicting the MSE, SSIM or VGG16 image difference from the distorted image alone while the reference is not observed.
This is enabled by two insights:
The first is to inject sufficiently many un-distorted natural image patches, which can be found in arbitrary amounts and are known to have no perceivable difference to themselves. 
This avoids false positives.
The second is to balance the learning, where it is carefully made sure that all image errors are equally likely, avoiding false negatives.
Surprisingly, we observe, that the resulting no-reference metric, subjectively, can even perform better than the reference-based one, as it had to become robust against mis-alignments.
We evaluate the effectiveness of our approach in an image-based rendering context, both quantitatively and qualitatively.
Finally, we demonstrate two applications which reduce light field capture time and provide guidance for interactive depth adjustment. 
\end{abstract}

\mysection{Introduction}{Introduction}
Computer vision or graphics experts easily recognize image artifacts that might be highly domain-specific.
An image-based rendering (IBR) specialist will quickly notice where depth estimation failed, where transparency was not handled or where a highlight did not move correctly.
Similarly, in computer graphics, artifacts resulting from Monte Carlo noise in image synthesis when producing a feature film, 
or shadow bias \cite{williams1978shadows} in a computer game are easily spotted by domain experts.The assessment typically is not limited to detection, but importantly includes judging magnitude as well as spatial locality.

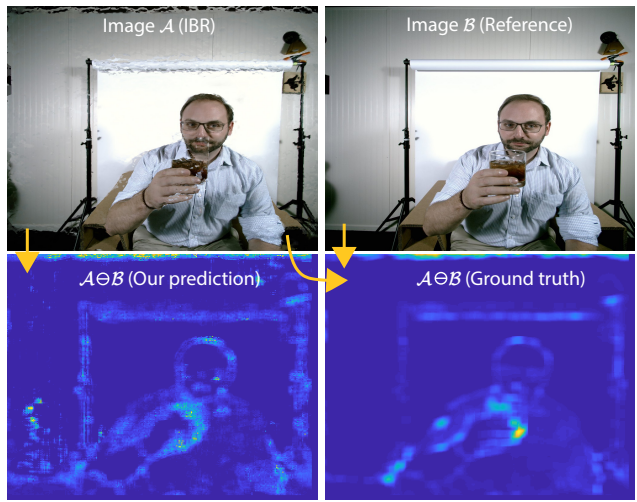
\myfigure{Teaser}{Given an image $\mathcal A$ \emph{(top left)} that is a version of a reference $\mathcal B$ \emph{(top right)} distorted by IBR artifacts, we predict their per-pixel difference map $\mathcal A\ominus\mathcal B$ \emph{(lower left)} without observing $\mathcal B$. The lower right shows the ground truth difference $\mathcal A\ominus\mathcal B$.
We here show MSE, but other metrics such as SSIM or VGG16 are also possible.
} 

The importance of interacting with errors can be seen from photographs with spatially annotated over- and under-expose artifacts, as done for instance by Henri Cartier-Bresson \cite{coleman2012lens}. Remarkably, all this is not achieved by comparing an image to a reference, but by experience and intuition built from knowing what natural images look like and how images with artifacts differ. Can we enable a machine to also perform such a task?

More formally, we face the challenge illustrated in \refFig{Teaser}.
Given an image $\mathcal A$ that is a distorted version of a reference $\mathcal B$ we wish to predict their difference $\mathcal A\ominus\mathcal B$ without access to $\mathcal B$.
The lower right image shows the ground truth metric response $\mathcal A\ominus\mathcal B$.
This metric could simply be the mean square error (MSE as used in \refFig{Teaser}), a more perceptual metric like SSIM \cite{wang2004image} or even VGG-16 activation differences that are effective as an image metric \cite{simonyan2014very,zhang2018unreasonable}.
More particularly, we go beyond the typical mean opinion scores \cite{talebi2018nima} given to uniform distortions such as noise or JPEG compression, and seek to produce localized distortion visibility maps without accessing the reference.

In this paper, we choose to study one specific form of artifacts that arise in image-based rendering (IBR) \cite{mcmillan1995plenoptic,gortler1996lumigraph}, in particular, when employed for novel-view synthesis from sparse light fields (LFs) \cite{levoy1996light}.
It is important in virtual reality and movie production where LFs are used to provide head motion parallax and special effects.
Moreover, having a localized error prediction is also important for quality control. In IBR, artifacts are very localized (\eg around certain depth edges) and creating opinion scoring or even spatio-angular annotated dataset of LF artifacts in a size sufficient for machine learning appears to be a daunting task.
Our method proceeds without all of this.

Addressing this challenge, we make use of convolutional neural networks.
We will show, how learning this mapping right away will result in many false positives or false negatives.
Instead, two important ingredients come together in our approach.
First, as the number of images containing artifacts is typically limited, we need to augment the training data with natural images that are free from artifacts.
Second, we propose a way to find the right balance between natural and distorted training data.

Not requiring a reference is useful whenever the original is inaccessible (lost, impossible to compute, unavailable, undefined).
Furthermore, we demonstrate one application of a non-reference metric in light field capturing.
We first capture a sparse light field, followed of by an interpolation of the intermediate views.
If our our metric indicate those intermediate views have errors, they views will be recaptured.
This allows acquiring higher-quality light field in much shorter time compared to dense LF capturing.

\mysection{Previous Work}{PreviousWork}

In this section, we discuss objective image quality metrics, with special emphasis on those that do not require the undistorted reference image.
Then, we briefly characterize IBR-specific artifacts, as well as metrics specialized in their detection, which is the key focus of this work.

\paragraph{Image metrics}
Some application and functions may require \emph{quality} while others need \emph{visibility} metrics \cite{Chandler2013}.

Image quality metrics (IQMs) evaluate the distortion magnitude and are typically trained on the mean-opinion score (MOS) data \cite{Sheikh06live,ponomarenko2009tid2008} that labels the entire image with as a single quality score.
The most commonly used IQMs such as PSNR, SSIM, MS-SSIM \cite{Wang06}, FSIM \cite{Zhang2011}, and CIELAB \cite{Zhang1997a} are \emph{full-reference} (FR) metrics that take as input the reference and distorted images, and compute local differences that are pooled into a global, single quality score.
Recently, it has been demonstrated that CNN-based FR-IQMs achieved best performance in predicting MOS data \cite{amirshahi_2016,Bosse2017}. 
Zhang et al. \cite{zhang2018unreasonable} employed crowdsourcing and created a large scale patch-based dataset in two perceptual experiments: (1) two-alternative forced choice (2AFC) on distortion strength, and (2) ``same/not same" near-threshold distortion visibility. They train different network architectures and report in each case a much better performance than traditional FR-IQMs in predicting their data from both experiments.

Visibility metrics (VMs) predict the distortion perceptibility for every pixel in the form of visibility maps.
VMs are specifically tuned for detecting near-threshold distortions, which is required in many graphics and vision applications that cannot tolerate any perceivable quality reduction and require local information on the distortion positions.
To decide on the visibility of such near-threshold distortions, models of human vision are often employed, where the most prominent FR-VMs examples include: VDM \cite{Lubin95}, VDP \cite{Daly1992}, and HDR-VDP-2 \cite{mantiuk11hdrvdp2}.
In the specific task of predicting selected rendering and compression artifacts, best performance has been achieved using machine learning \cite{vcadik_2013} and CNN-based techniques \cite{wolski2018dataset,patney2018detetcing}. 

\paragraph{No-reference metrics}
In this work, we focus on the VMs due to the locality of their prediction, but we are specifically interested in more challenging \emph{no-reference} setup, where the reference image is not available.
We discuss the most successful and recent NR-IQMs that rely on machine learning techniques, and we also refer the interested reader to more comprehensive metric surveys in \cite{Chandler2013,Kim2017a}.
Early machine learning techniques employed predefined features such as SIFT and HOG \cite{Narwaria2010,moorthy10twostep,saad12,tang11learning}, and measured their distortions with respect  to natural image statistics \cite{Wang06}. 
Recently, CNN architectures are applied to such feature learning as well as the MOS regression at the same time \cite{bianco_2016,kang_2014,Bosse2017,talebi2018nima}.
To compensate for a low number of MOS-labeled images, such solutions typically rely on patches, where they assign the same MOS score for all patches that belong to a given image \cite{Kim2017a}.
Such practice is justified for specific classes of distortions that affect the whole image uniformly, which might be the case for certain types of image noise or compression artifacts, but might confuse the network in case of localized distortions such as those occurring in IBR.

To compensate for the lack of true local reference images, Bosse~\etal \cite{Bosse2017} learn the importance of local patches, but their key motivation is not in deriving the localized VM, but rather in estimating relative patch weights in the aggregated MOS rating.
Lin and Wang~\cite{Lin2018Hallucinated} employ a quality-aware generative network to hallucinate the reference image, which by employing adversarial learning is further refined by an IQM-discriminator that is trained on ground truth references. Their hallucination-guided quality regression network is fed with the difference between the hallucinated and distorted images, as well as the distorted image itself to predict the MOS value. 
The quality-aware generative network, hallucination-guided quality regression network, and the IQM-discriminator are jointly optimized in an end-to-end manner.
Kim and Lee~\cite{Kim2017} apply state-of-the-art FR-IQMs such as SSIM to generate proxy scores on patches as the ground truth to pre-train the model and then fine-tune their target NR-IQM. 
At intermediate stages the regression network considers mean values and the standard deviations of per-patch 100-element feature vectors which are then pooled to a per-image quality score. 

In this work, we also employ state-of-the-art FR-IQMs to perform an initial per-patch distortion annotation, and strike the required balance between different error magnitudes in the training data, which is essential for meaningful training and shift-invariant properties of our NR-VM.

The research on NR-VMs is extremely sparse, presumably due to limited access to locally labeled images \cite{herzog_2012,vcadik_2013,wolski2018dataset}.
A notable exception is the work of Herzog et al.~\cite{herzog_2012} who employs support vector machine (SVM) to predict per-pixel distortions for selected rendering artifacts (they do not consider IBR) and achieve performance comparable to FR-VMs. 
Here, we demonstrate that time-consuming manual per-pixel distortion labeling is not strictly required.

In cases where training data is both easy to produce--such as uniform distortions like noise, JPEG, etc.--and no perceptual calibration is required, supervised training has been employed to detect aliasing artifacts \cite{patney2018detetcing}.
Our work differs, as we only have very limited training data available, both because only very few ground truth images are available for IBR and we need perceptual calibration.
Learning from little data is part of our balancing contribution.

Vogels and colleagues \cite{vogels2018dneoising} have proposed a method to denoise path traced images.
To steer the amount of denoising, they also trained a neural network to predict distortion in terms of MC variance, which is as unknown as the pixel value to be MC-estimated itself.
Interestingly, in both their work and ours, a NR metric is used to steer adaptation: for them it is a denoising algorithm; for us, one application is controlling capture hardware.
Their task is different as they predict SSIM error from a pair of images, where one is noisy and the other is denoised. This restricts the distortions to the difference between denoised and reference, which are smaller than IBR artifacts and also does not need to be perceptually calibrated.
The fact that images with MC noise can be generated in arbitrary amounts also underlines what is the focus of our work: coping with limited training data.

\paragraph{Image-based rendering} for structured or unstructured light fields (LFs) of real-world scenes involves a number of computational steps such as: depth reconstruction, neighboring view-image warping, warped view-image blending, and disocclusion hole in-painting. Each of these steps is prone to inaccuracies that manifest themselves as IBR-specific artifacts such as object shifting (incorrect depth), crumbling, distorted edges (depth discontinuities, \eg due to compression), popping (fluctuations in depth), ghosting (depth inaccuracy, view blending), stretching, blurry or black regions (in-painting) \cite{Tian2018}. Specialized IBR quality metrics often rely on leaving one view out as the reference \cite{Waechter2017,Conze2012,Solh2011,Bosc2011} or searching for matching image blocks after their registration  \cite{Battisti2015,gu2017model}, and then employing customized FR-IQMs. NR-IQMs typically focus on detecting selected distortion types such as blurring and ghosting \cite{Berger2010}, ghosting and popping \cite{Guthe2016}, blurring, stretching and black holes \cite{Tian2018}, and aggregation into one final scalar score. Perceptual experiments have been performed to understand how the observers rate the severity of different artifacts as a function of rendering parameters such as the number of blended views and viewing angles \cite{vangorp2011perception}. A skillful pre-processing of depth (\eg depth blurring in uncertain regions) and choice of particular algorithmic solutions can substantially suppress artifacts \cite{hedman2016scalablerendering,serrano2019motion}, eventually using a neural network trained to predict blending weights to combine the warped images\cite{hedman2018blending}. More objectionable distortion types can be traded-off with those that are more visually appealing (\eg blurry depth that is more consistent but further from the ground truth).
Instead of focusing on selected distortion types, Ling et al \cite{ling2018learn} proposes to learn a dictionary based on manually labeled data.
The features extracted from an image allows to predict a MOS value using support vector machine regression.
As data labeling can be time consuming, as Ling et al. \cite{ling2019gans} create artificial training data that aims to simulate occlusion problems.
A Generative Adversarial Network (GAN) discriminator \cite{goodfellow2014generative}, targeted to identify in-painted image regions, is used to predict a quality score.

All the discussed work on IBR quality evaluation essentially focuses on providing a single score per-image, which then also serves as a metric for performance evaluation. While some FR-IQMs generate viable per-pixel VMs at intermediate stages \cite{Conze2012,Solh2011}, their accuracy is not formally evaluated. The same holds for the NR-IQM \cite{ling2018learn}. Our work hence differs from all previous work by pursing the NR-VM setup to detect local IBR distortions using CNN-based techniques. 

\mysection{Learning a No-reference Metric}{OurApproach}

\paragraph{Overview}
Test-time input to our method is a single distorted RGB image $\mathcal A$.
While our distortions are always IBR artifacts resulting from a specific depth reconstruction and specific IBR method, the interna of how this image is generated (\eg the depth map) are transparent, and we only need access to the result.
Withheld is the reference RGB image $\mathcal B$.
In the case of IBR, such a distorted-undistorted pair is typically produced by rendering a known image from other known views.

Output of our proposed method is a single-channel (scalar) image that predicts a given difference metric response $\mathcal A\ominus\mathcal B$, where the $\ominus$ operator depends on the choice of the specific metric, \eg MSE, SSIM \cite{wang2004image}, or VGG16 \cite{simonyan2014very}.
High values are produced where the images are different and small values where they are similar.
This output is accurate, if it has little false positives or negatives.
False positives correspond to predicting a perceived difference where there are no artifacts and false negatives correspond to visible artifacts the metric fails to report.

Note that two forms of approximations are made here: the first is the error that the metric itself makes when comparing two images relative to human judgment.
The second is the error that our method has, with respect to a prediction.
Ultimately, our method is a prediction of a prediction, but surprisingly can perform better than one prediction alone. 

\mysubsection{Training data}{TrainingData}
Our training data comprises existing metric responses $\mathcal A\ominus\mathcal B$ to the distorted image $\mathcal A$ and the clean reference image $\mathcal B$.
Strictly speaking, learning does not even observe the reference image $\mathcal B$, but in practice, it is required to compute the metric response $\mathcal A\ominus\mathcal B$.

For creating our training dataset, we used captured LF images of 42 different scenes, which come from the Stanford LF repository \cite{stanfordLF}, the Fraunhofer IIS light field dataset \cite{dabala2016lightfield}, Google Research work \cite{Soft3DReconstruction}, and Technicolor \cite{Sabater2017} as well as from our own captured images.
All 4D LF datasets comprise conventional 2D images in a resolution up to 2k$\times$2k, taken from a range of sparse view points, such as in a 3$\times 3$ camera array with known camera positions.
For each LF view point, we first estimate the depth using a light field depth estimation technique \cite{dabala2016lightfield} and then warp \cite{mark1997post} the image into all other views.
For each LF, we use the four corner views to generate novel-view images at the positions of the remaining views.
Each warped view corresponds to one original view, and we compute the response of a full-reference metric to this pair.
With approx.\ 9 views per LF and 42 LFs in total, this amounts to only 210 unique images, \ie a comparatively low number for a training task.

We use six scenes for testing and the rest for training.
The same split is also applied later for the user study.
Our test scenes are totally different from the training scenes, which is important as the number of scenes in the training set is small and generalization across them is an additional challenge.

The natural images used in our training and test dataset are sourced from the Inria Holidays image dataset \cite{jegou2008hamming}  which have a comparable resolution to our LF images.

Our method is independent of the actual underlying metric $\ominus$ we predict.
We will denote this response neutrally as $\mathcal A\ominus\mathcal B$.
We explored three metrics: MSE, SSIM and VGG16.
MSE is defined as the average per-pixel RGB difference vector length squared.
The SSIM metric is using the original implementation \cite{wang2004image}.
VGG16 \cite{zhang2018unreasonable} transforms both $\mathcal A$ and $\mathcal B$ into the VGG16 feature space and picks the activations at layer five, which is 512-dimensional.
The $L_2$ difference of these two vectors is used as the metric response.
For each metric, we normalize the 95th percentile of their responses across the training dataset to fall between 0 and 1.

\mysubsection{Architecture}{Architecture}
We use a simple encode $P$ \cite{ronneberger2015unet} that has learnable parameters $\Theta$ and predicts the error map $P(\mathcal A | \Theta)$ by observing $\mathcal A$ (\refFig{Architecture}).

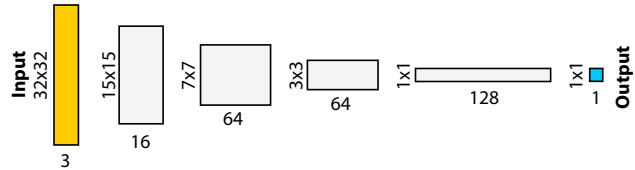
\myfigure{Architecture}{Our architecture consumes $32\times32$ patches, (yellow left), and applies a cascade of $3\times3$ convolutions, followed by non-linearities (ReLU).
Spatial resolution is reduced \emph{(height)} and feature count increases \emph{(width)} before a final prediction of the metric response is produced \emph{(blue, right)}.}
The network comprises 5 layers ($32\times 32$ patch size) with the  total number of $|\Theta|=175,537$ learnable parameters and is trained on all patches of the training set in a sliding window fashion.

The loss is the $L_1$ error of the predicted metric response, so $||P(\mathcal A| \Theta)-(\mathcal A\ominus \mathcal B)||_1$.
Note that the loss is always $L_1$, while the metric can be the $L$-norm-like MSE as well as SSIM or VGG16.

\paragraph{Balancing}
We have explained why, and will see from the ablation study, that it is important to have natural patches, but the question is how many.
If we take an unlimited number, the metric prediction simply always returns zero, because natural patches have no error to themselves.

\myfigure{PatchSampling}{
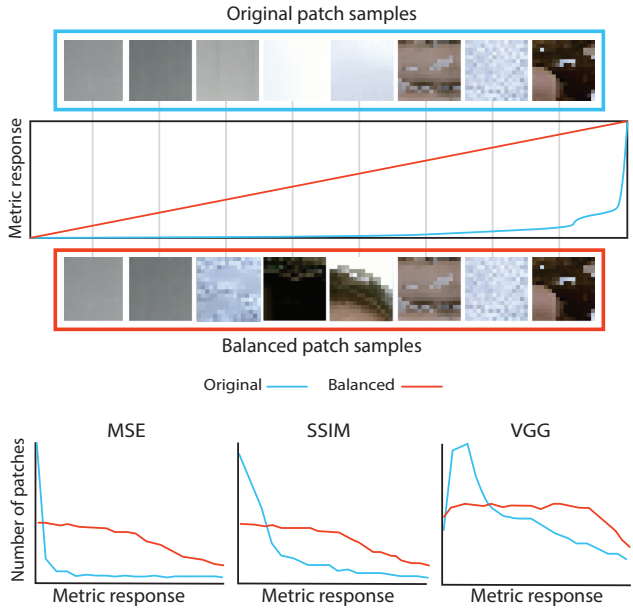
When sampling uniformly from IBR patches the error distribution 
is skewed towards low errors \emph{(blue)}.
Our balancing \emph{(red)} adjusts the samples to have a uniform range of errors.
The
three lower plots show the actual distribution before and after balancing
for different metric responses.
}

Our solution is to start with a half-half mix of distorted and clean patches.
Regrettably, many of the distorted patches, which make 50\,\% of the total, also have small errors that are close to zero.
These patches are exactly those for which IBR was successful, \ie did not have any artifacts.
Depending on the metric, this imbalance can be very strong, and in particular for MSE, it is extremely heavy-tailed (\refFig{PatchSampling}).
To address this, we balance the error distribution for the distorted half when creating the training data as follows:
First, we sort all patches by their metric response into a priority queue.
Then, we uniformly random-sample  the range from zero to the 95th percentile of the metric response distribution.
For every sample $i$ with value $\xi_i$, we find the patch $j$ with the most similar metric response $d_i$ and remove it from the queue and add it to the training dataset.
When the minimum difference $\xi_i-d_j$ is larger than a threshold $\epsilon$, we reject the sample.
This is repeated until a target patch count, such as 250\,k, is reached.

\mysection{Evaluation}{Evaluation}

\mysubsection{Methods}{Methods}

\paragraph{Training Strategies}
We compare three different strategies for training.
The first is ours, the other two are ablations.
\textsc{Full} is our complete method involving 50\,\% natural patches and a balancing of the other 50\,\% as described in \refSec{Architecture}.
\textsc{NoBalance} is realized by a similar 50/50-split, but we train on all distorted patches without the balancing.
\textsc{NoNatural} adapts the balancing to take 100\,\% of the patches coming from IBR without adding the natural patches as described in \refSec{Architecture}.
All training sets, albeit processed differently, have the same size of ca.~.5\,M patches.

\paragraph{Error}
As we predict metric responses, our error is the same as the loss, the absolute difference between the ground truth metric response and our prediction of that response.
As these errors also come in arbitrarily different scales for different metrics, we normalize them per metric by dividing by the global 95th percentile of the GT metric response across the balanced training dataset.

We additionally report errors in metric prediction errors for a split subsets to understand the false/true-positive and false/true-negative tendency.
In \textsc{All}, we compute the error for the whole test dataset. 
Additionally, we consider two subsets of the test dataset.
The first subset is \textsc{Clean}, which includes only natural patches.
The second one is \textsc{Distorted} that contains only IBR patches, including those that might also come out with very low or even with no error.
Please note that this is a partitioning of the test set, and not of the training set.

\mysubsection{Quantitative results}{Quantitative}
In this section, we discuss both the means and full error distributions of all training strategies for different partitions and different metrics.

\begin{table}[h]
\setlength{\tabcolsep}{2pt}
\caption{
Error of the metric predictions on the test data for different variants of our algorithms and different partitions (\textsc{All/Clean/Distorted}) of the training data \emph{(columns)} on different metrics \emph{(rows)}.
Winners per-partition  are marked bold.
}
\label{tbl:Main}
\center
\begin{tabular}{l rrr rrr rrr}
\multicolumn{1}{c}{\multirowcell{2}{Metric}}
&\multicolumn{3}{c}{\textsc{Full}}
&\multicolumn{3}{c}{\textsc{NoNatural}}
&\multicolumn{3}{c}{\textsc{NoBalance}}
\\
\cmidrule(lr){2-4}
\cmidrule(lr){5-7}
\cmidrule(lr){8-10}
&\multicolumn{1}{c}{\small\textsc{All}}
&\multicolumn{1}{c}{\small\textsc{Cle.}}
&\multicolumn{1}{c}{\small\textsc{Dist.}}
&\multicolumn{1}{c}{\small\textsc{All}}
&\multicolumn{1}{c}{\small\textsc{Cle.}}
&\multicolumn{1}{c}{\small\textsc{Dist.}}
&\multicolumn{1}{c}{\small\textsc{All}}
&\multicolumn{1}{c}{\small\textsc{Cle.}}
&\multicolumn{1}{c}{\small\textsc{Dist.}}
\\
\toprule
\multirowcell{1}{MSE}
&\textbf{.098}
&{.006}
&{.189}
&{.137}
&{.092}
&\textbf{.182}
&{.102}
&\textbf{.003}
&{.201}
\\
\multirowcell{1}{SSIM}
&\textbf{.078}
&{.013}
&{.143}
&{.143}
&{.159}
&\textbf{.127}
&{.080}
&\textbf{.012}
&{.149}
\\
\multirowcell{1}{VGG}
&\textbf{.085}
&\textbf{.006}
&{.165}
&{.207}
&{.293}
&\textbf{.121}
&{.092}
&{.008}
&{.176}
\\
\bottomrule
\end{tabular}
\end{table}

\paragraph{Means}
The means of all methods are compared in \refTbl{Main}.
We see that our method (\textsc{Full}) has the smallest error across different metrics compared to both other variants (bold in column \textsc{All}).

In detail, when we look into the partitioning, we find that for the \textsc{Distorted} partition, the \textsc{NoNatural} strategy performs best.
This is expected as training  is done with all distorted patches which comprise the maximal variety of distortion.
This makes the resulting metric sensitive for all kinds of distortions.
As a result, the probability of false negatives, \ie claiming patches with an error to be fine, becomes low.

We also find, that for the \textsc{Clean} partition, the \textsc{NoBalance} strategy performs best.
This also is expected as in the training, 50\,\% of data comprises natural (undistorted) patches, and due to the \textsc{NoBalance} strategy, small errors dominate in the distorted patches. This makes the resulting metric particularly sensitive for near-threshold distortions.
In this case, the probability of false positives, \ie reporting a high metric response for no-error patches, is low.

All statements are true (significant, $p<.01$, $t$-test after testing for Gaussianity) across all metrics, indicating that the \textsc{Full} approach is independent of the underlying metric.
A positive exception is VGG, where the \textsc{Full} approach even performs better than \textsc{NoBalance} on the \textsc{Clean} partition.

\paragraph{Distributions}
In \refFig{Analysis}, we show the distribution of errors for different metric predictions (top) and the correlation of the prediction error and metric response (bottom).
In each plot, colors encode the variants of our approach (\textsc{NoNatural}, \textsc{NoBalance}, \textsc{Full}).

\myfigure{Analysis}{
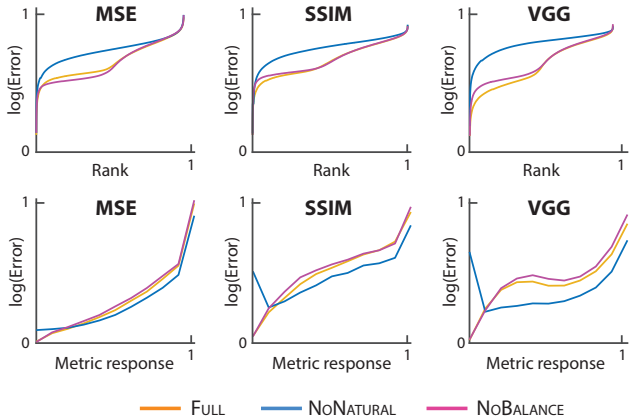
Analysis of metric prediction error, for different metrics and variants of our method.
The top plots show sorted error distributions.
The bottom row plots show the correlation of metric response and metric prediction error.
All vertical axis are log scale.
}

Each plot in the first row of \refFig{Analysis} shows the sorted error of our metric prediction in ascending order.
We see that across the entire range, with the exception of MSE prediction for low errors; the \textsc{Full} approach performs better than other variants.
This indicates that the mean is a good characterization of the performance.
In all cases, we noticed a sudden increase in the error that occurs around 50\,\% of the population, \ie the error for the first half of the population seems to follow a different trend than the second half.
We hypothesize that, these are the patches where reference and input are (partially) not aligned, which make up roughly 50\,\% of the population as well.
Unfortunately, there is no way to tell apart a misaligned patch that is judged by FR metrics as different with respect to a displaced reference.
Hence, large errors are expected to become undetectable at some error level.
The exception is the regime in MSE where the \textsc{Full} approach is worse on low errors and slightly better on high errors, while it performs best on average in (\refTbl{Main}).
This can be difficult to comprehend due to the log scale of the vertical axis.

Each plot in the second row in \refFig{Analysis} shows the error of our prediction on the vertical axis and the metric response on the horizontal axis as a connected scatter plot.
We can see that the plots are in accordance with  \refTbl{Main}: The \textsc{NoNatural} method which performs best in predicting high metric responses, has a high error on patches with small metric response (false positives).
Symmetrically, the \textsc{NoBalance} method which is the best at predicting low metric responses, produces high errors on patches with high metric response (false negatives).
\textsc{Full} method is always a bit worse than one other method in one region (except at the unique point where both cross), but on average performs best overall.

\mysubsection{Qualitative results}{Qualitative}

\mycfigure{MainResults}{
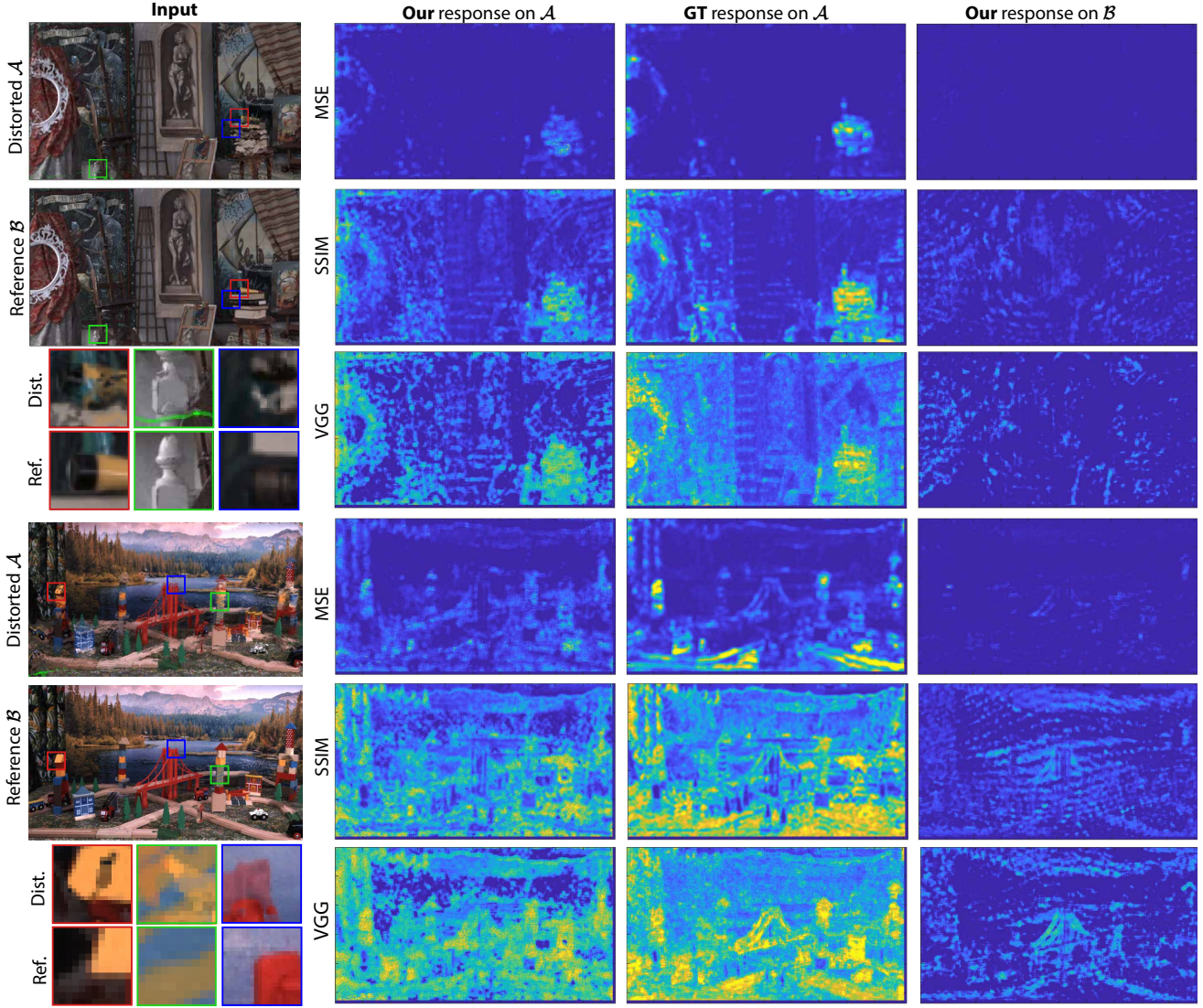
Comparing the response to a pair of an image $\mathcal A$ and its distorted version $\mathcal B$ \emph{(first column)}.
Our response \emph{(second column)} is similar to the ground truth \emph{(third column)}.
When executed on the clean reference \emph{(fourth column)}, only very few false positives are reported.
}

\paragraph{Example metric outputs}
\refFig{MainResults} shows an analysis of the response of all metrics to two different LFs from the test set.
The first column shows the distorted input $\mathcal A$ in the top, below the hidden reference $\mathcal B$ and below this three insets from both.
The second column shows our predicted response $\mathcal A \ominus \mathcal B$ for different metrics: MSE on top, followed by SSIM and VGG.
A false color coding, where cold colors indicate a low response and warm colors indicate a high response, is used.
The third column shows the GT response for the same.
It is evident that there is a similarity between our prediction and the ground truth. We slightly err towards conservative, \ie miss a few errors.
How some of these errors are only false findings, \ie a limitation of the metrics, becomes apparent from the user study to follow.

The last column shows a sanity check where we put the hidden reference image $\mathcal B$ into our metric.
The hidden reference obviously does not contain any error, and consequently reporting one is a false positive.
We see, that our image has a responses in areas that are correct but look like IBR artifacts, but in most areas has no response.
In summary, this indicates that we localize and scale errors to a hidden reference in images with artifacts, while avoiding to produce a signal when facing clean images. 
It might appear that MSE has less false positives than SSIM or VGG when inspecting the last column; simply more deep blue, very close to perfect in the first row.
However, such a trend is not supported by the numbers  in \refTbl{Main} or the plots in \refFig{Analysis}.
The true reason for this impression might be that the SSIM and VGG response simply have a larger receptive field per-se: MSE is per-pixel while VGG is affected by up to $32\times 32$ pixels.
Even the ground truth response is more dense (less deep blue).
Consequently the metric prediction, in case of error, also makes spatially more extended, more dense, mistakes.

\mycfigure{invariance}{
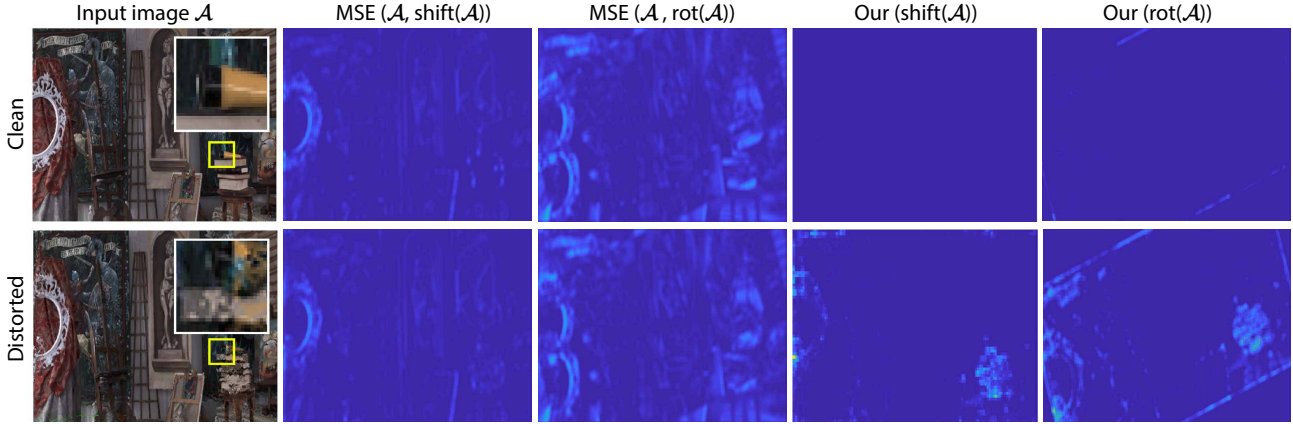
Transform-invariance of our approach:
When computing the distance between a clean input image $\mathcal A$ \emph{(first column, first row)} and a misaligned reference $\mathcal B$ (not shown here, 20-px shifted or 20 degrees rotated copy of $\mathcal A$), a common metric such as MSE will show a strong response \emph{(first row, second and third columns)}.
Such a response is numerically correct, but far from human assessment, which would be more similar to our response \emph{(first row, fourth and fifth columns)}.
Symmetrically, repeating the experiment on a distorted input, our approach correctly localizes the distortions around the books \emph{(inset)} as if the reference had been aligned.
}

\paragraph{Transformation-invariance}
Surprisingly, results produced by our approach can turn out to be better than their own supervision, as our method is forced to come up with strategies to detect problems without seeing the reference.
This makes it immune to a common issue of many image metrics: misalignment \cite{Kellnhofer2016jei}.
Even a simple shift in image content will result in many false positives for classic metrics (\refFig{invariance}).
An image that has merely been shifted is reported to be very different from a reference by all the metrics used for our supervision;
however, it shows less differences in case we add IBR artifacts to it.
In contrast, our method does not care about transformation, but when IBR artifacts are added, they are detected.
As our proposed method is oblivious to the ground truth, it is not subject to such a misconception.
While not quantifiable, the result is arguably more similar to human judgment, as indicated by the user experiment in the next subsection.

\mysubsection{User study}{Study}
We have conducted a user experiment to validate that our predicted metric responses spatially correlate with the visibility of artifacts to human subjects.
We quantify the human responses by means of per-pixel annotations, which are painted on top of images showing IBR artifacts.
Note that no user responses was used for training.

\myfigure{UserStudy}{
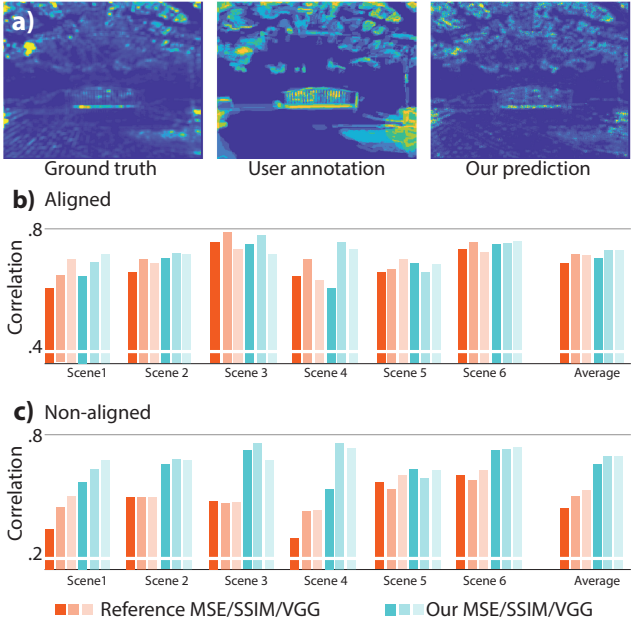
Exemplary user study result \emph{(a)}.
Correlation (significant, $p<.001$) of MSE/SSIM/VGG and user responses (red) compared to our predictions of the three metrics (blue) for different scenes and as an average across scenes to the right \emph{(b)}.
We can see that in the non-aligned conditions, these differences get stronger \emph{(c)}.
}

\paragraph{Methods}
Na\"ive users were asked to use a binary painting interface to mark errors in a rendered image for each of the six LFs of our test dataset in an open-ended session that took 15 minutes on average.
We average the binary response into a continuous fraction (percentage) of users that detected the location of the artifacts.

\paragraph{Analysis}
Asking $N=10$ users, we find the correlation (Pearson linear correlation $R$, higher values are better; statements highly significant as the correlation is computed on a high number of image pixels) reported in \refFig{UserStudy}-b.
We see that for many scenes as well as for the average across scenes, our method has a higher correlation with user annotation than the metric it was supervised on.
We hypothesize, that this is due to the fact that our network had learned to become independent of a reference, a similar robustness that the HVS employs.
There is no clear trend on which of our metric response predictions  correlates the most with the user annotations.
The differences between scenes, however, seem more pronounced.

When repeating the experiment with a non-aligned reference (shifted a mere 20~px to the right), we find the correlations reported in \refFig{UserStudy}-c.
We see that our correlation even improves in this condition(our metric shows higher correlations for all metrics across different scenes), showing we are more robust to alignment issues when predicting user responses.

\paragraph{Perceptualization}
Finally, we computed a linear correlation $R$ by fitting a model $x_i=a\cdot y_i+b$, where $x_i$ is the user response and $y_i$ is our prediction of the metric response for pixel $i$.
This allows a ``perceptualization'' of our metrics response.
Fitting multiple models $a,b$ in a leave-one-out protocol to 5 of our 6 scenes produces an average error of $.05/ .04/ .02$ for MSE/SSIM/VGG respectively, indicating that this perceptualization generalizes to some extent.

\myfigure{Adaptive}{
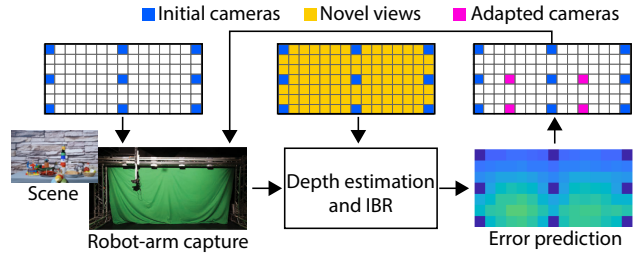
Proposed pipeline for adaptive LF sampling by bounding the reconstruction error predicted by our no-reference metric.
}

\mysubsection{Other architectures}{OtherArchitectures}
We also explored using other architectures with or without balancing.
A simple solution would be to use a supervised image translation network such as Pix2Pix \cite{pix2pix2016} to map from entire IBR images to the metric response.
Unfortunately, training these on our data  converges to a flat response of zero, as artifacts are too rare and subtle to be picked without the balancing we suggest.
Future work could investigate combining our balancing with other architectures.

\mysubsection{Supplementary materials}{Online}
Ground-truth responses of all metrics and our predictions for all input images, for all variants of the algorithm, as well as all user study annotations can be explored in an interactive web application in the supplementary materials.

\mysection{Applications}{Adaptive}
We will now demonstrate two practical applications of a NR-IQM in light field production.
The first is accelerating automated adaptive LF capture (\refSec{AdaptiveCaptureApplication}), the second employs our NR-IQM as a feedback in an interactive depth manipulation system (\refSec{InteractiveDepthAdjustment}).

\mysubsection{Adaptive light field capturing}{AdaptiveCaptureApplication}
Capturing a dense set of input view images results in a high-quality reconstruction but remains a time-consuming process or may require a bulky setup.
Our main observation is that not all input view images contribute equally to the reconstruction of novel-view images.
Our metric helps identifying and capturing these.

Images from views dominated by planar diffuse surfaces can reliably be predicted from images taken from other views showing this very same surface.
Hence, dense capturing from these views is needed and thus not efficient.

In contrast, occlusions and specularity can be more challenging, because it must be ensured that each scene element is visible in at least two camera views (when using multi-view stereo, as we do) to compute depth.
Sparse capturing from these views would sacrifice the reconstruction quality.

To both of these ends, we propose an adaptive capturing mechanism as it illustrated in \refFig{Adaptive} to capture an image for a view only if it cannot be extrapolated from other views.

\mysubsubsection{Setup}{Setup}
We study adaptive capturing by means of a large-scale translation stage equipped with a digital camera.
The position of the camera can be controlled with a precision of $80\,\mu m$ in horizontal and $50\,\mu m$ in vertical direction.
This allows for very dense capturing of the scene.
While this takes long to capture, it serves as a unique baseline to our study where we can compare our prediction of an error to the actual error present.

\mysubsubsection{Procedure}{Procedure}
We first capture a sparse set of images and estimate the depth maps for each view.
Then, we use DIBR to render a set of intermediate-views and compute the reconstruction error for each rendered view.
All pixels are simply averaged in each view image, producing a single scalar value.
The capturing grid is then subdivided into smaller regions where average predicted reconstruction errors is larger than a given threshold.
This process is repeated until a desired quality is achieved.
By this approach, the number of captured views can be substantially reduced, and we only need to capture images at locations where reconstruction is poor.

\myfigure{comparison}{
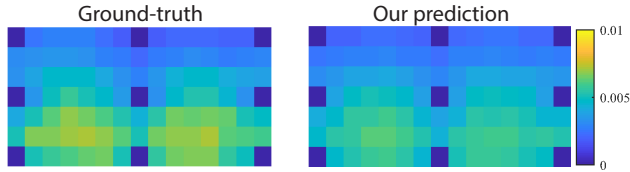
Reconstruction error of intermediate novel views. \emph{Left}: Ground truth MSE values, \emph{right}:  Our network MSE prediction.  
}

Predicting the reconstruction error of novel view  is the key to make such an approach work.  Classic full-reference image quality metrics require a dense capture to provide reference images to compute the error, which is not practical as our goal is to reduce the number of captured images in the first place.
In contrast, our proposed no-reference metric can measure the error in the novel view images without providing their reference images, resulting in an efficient approach.

\mycfigure{Sampling}{
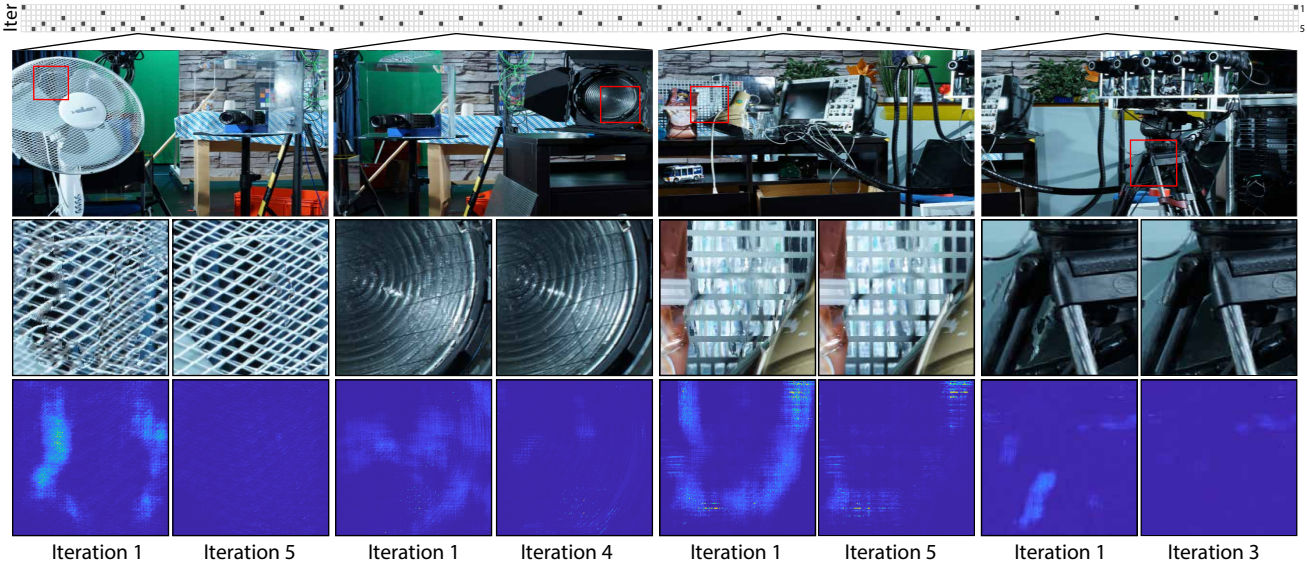
Adaptive panoramic light field capturing: The top row shows a grid indicating the camera placement at different iterations.
The second row shows the selected rendered views based on the key frames that are captured. The insets in the third row show the marked patches from the rendered views in the first iteration and in the iteration that a desired quality is achieved. In the fourth row, we also show our network predictions for the corresponding patches in each iteration.
}

\mysubsubsection{Evaluation}{Evaluation2}
To evaluate effectiveness of our metric in this application, we
simulate capturing two LFs, adapted according to the MSE metric.

\paragraph{Array}
We captured an array of 7$\times 15$ images for the scene shown in \refFig{Adaptive}\,(left).
In \refFig{comparison} we show the ground truth MSE (left) and our network prediction (right), where each grid element denotes a camera position.
The dark blue grid elements indicate the camera positions where actual key frames were captured, while rendering has been performed for all remaining intermediate positions.

As we can see, the distribution of reconstruction error as predicted by our metric correlates well with the ground truth. \refFig{Adaptive}\,(right) shows new camera locations that are required to reduce the true average reconstruction error below .004.

\paragraph{Panoramic}
We also demonstrate the potential benefit of our approach for an efficient panoramic (\ie one-dimensional, linear) light field capturing. 
As it is shown in \refFig{Sampling}, depending on the scene content, not all regions in the scene require equally dense camera placement.
Our metric successfully guides the capturing setup to take more photos in the regions with thin structures, substantial disocclusions or specularites where accurate reconstruction is highly challenging.
Overall, capturing 76 instead of 720 images -- a sparsity of 10.5\,\% -- reduces the total capture time from 59 minutes to 4.9 minutes, \ie by 91\,\%.

\mysubsection{Interactive depth adjustments}{InteractiveDepthAdjustment}
Long acquisition times involved in capturing dense light fields make it a tedious and impractical task for some application fields.
One of such fields is movie production, where the presence of highly dynamic scenes and time pressure discourages the use of dense light fields, and in such cases, only sparse light field capture using video camera arrays is seen as a convenient solution.

Unfortunately, automatic error-free light field reconstruction from a sparse capture is still an unsolved problem.
To this end, there are ongoing research efforts to address the challenges such as the estimation of disparity in the presence of homogeneous areas, repetitive structures, fine-grained objects, or specularities.
In such cases, interactive disparity estimation improvement seems to be the most promising solution to achieve a high-quality view rendering \cite{wildeboer2011semi,kap2015apparatus,lin2012interactive,cao2011semi}.
However, this requires detecting possible view rendering artifacts as fast as possible to reduce the post-processing time.
As shown in the right-most image of the second row in \refFig{Sampling}, spotting an artifact is not a trivial task and sometimes requires carefully scanning the view rendering result.
Our quality estimation metric can significantly simplify this process by allowing the automatic analysis of several novel rendered views.
By observing the predicted visibility map, which identifies the local distortions, the user can quickly spot the problematic regions.
Using a post-production software suite \footnote{\url{https://www.iis.fraunhofer.de/realception}} to perform an interactive view rendering with only a small subset of cameras allows detecting the captured view responsible for the error.
The inspection of the corresponding disparity map followed by an approach similar to \cite{wildeboer2011semi,kap2015apparatus} finally allows fixing the view rendering error.
This is achieved by manual creation of a geometry proxy in 3D space for objects whose disparity map could not be computed automatically.
The proxy is then used to bound the admissible depth values for a subsequent disparity estimation.

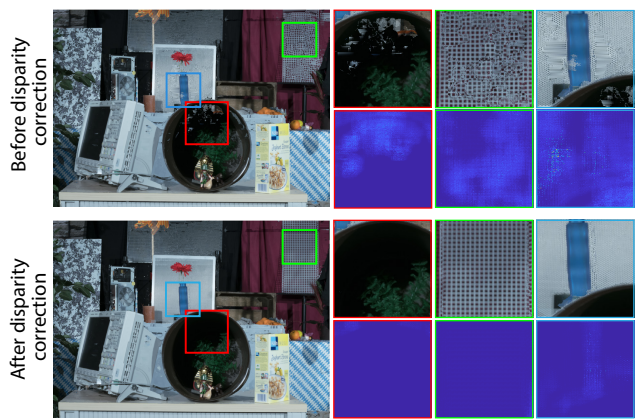
\myfigure{interactive}{Interactive depth adjustment. The marked patches are showing the regions in the rendered view where our method predicts the MSE (top) and the bottom row shows the corresponding patches after applying our manual disparity refinement.}

The results of this procedure are illustrated in \refFig{interactive}.
The contained repetitive structures are very challenging for automatic disparity estimation and consequently lead to many view rendering artifacts as clearly indicated by the depicted error map.
For solving these issues, a user has added proxy-based disparity constraints for the waste basket (and the contained figurine), the grid structure behind the flower, and the grid structure in the upper right corner of the image.
By these means, a much better view rendering could be achieved as shown in \refFig{interactive}.
Our metric has reduced the time required to find those reconstruction errors, leaving more time to a user to correct them.

\mysection{Conclusion}{Conclusion}
We have demonstrated that with properly adjusted training data (prioritization and natural supervision), a CNN can learn how to predict the difference of an image to a hidden reference.
Our approach is independent of the metric used and we have shown MSE, SSIM and VGG prediction.
Other metrics such as HDR-VDP-2 \cite{mantiuk11hdrvdp2} or the CNN-based metric of Wolski \etal\cite{wolski2018dataset} would likely be predictable in a similar fashion.

Such a metric can be applied for several applications.
As demonstrated this includes adaptive light field sampling of complex scenes and interactive depth editing.
Moreover, since in contrast to any existing non-reference metric, our approach provides a predicted error map, this opens the potential for many novel applications such as interactive or automatic view rendering error correction.

In future work, we would like to overcome the limitations of the paired input, eventually using an adversarial \cite{goodfellow2014generative} design, and learn the prediction only from pairs and without the metric, or only from pairs of undistorted-metric or distorted-metric.

{
\small
\noindent
\paragraph{Acknowledgements}
This work was partly supported by the Fraunhofer-Max Planck cooperation program within the framework of the German pact for research and innovation (PFI) and a Google AR/VR Research Award.
}


\end{document}